\documentclass[letterpaper,12pt,oneside]{article}

\usepackage[utf8]{inputenc}
\usepackage{mathptmx}
\usepackage{lipsum} 
\usepackage{titlesec}
\usepackage{ragged2e}
\usepackage{fancyhdr}
\usepackage[all]{nowidow}
\usepackage{url}
\usepackage{tocloft}
\usepackage[headheight=110pt, top=1in, bottom=1.5in, left=1.2in, right=1in]{geometry}

\usepackage{filecontents}
\usepackage{verbatim}
\usepackage{comment}
\usepackage{hhline}
\usepackage{amssymb}
\usepackage{amsmath}
\usepackage{makecell}
\usepackage{multirow}
\usepackage{colortbl}
\usepackage{graphicx}
\usepackage{subcaption}
\usepackage{mwe}
\usepackage{mathpazo} 
\usepackage{longtable}
\usepackage{lmodern}
\usepackage{mathtools}
\usepackage{etoolbox}
\usepackage{enumitem}
\usepackage[acronym]{glossaries}
\usepackage{lscape}
\usepackage{afterpage}
\usepackage{setspace}
\usepackage{pdflscape}
\usepackage{rotating}
\usepackage{etoolbox}
\usepackage{changepage}
\usepackage{authblk} 

\usepackage[authoryear]{natbib}        
\usepackage[colorlinks=true, citecolor=darkred, urlcolor=darkred, linkcolor=darkred]{hyperref}
\usepackage{journal_abbreviations}

\newcommand{\ngas}{$n_{\rm gas}$}
\newcommand{\Tgas}{$T_{\rm gas}$}
\newcommand{\Zgas}{$Z_{\rm gas}$}
\newcommand{\hmol}{${\rm H}_{2}$}
\newcommand{\barra}{$-$}
\newcommand{\mycol}[1]{\begin{tabular}{c} #1 \end{tabular}}

\definecolor{boh}{rgb}{0.0, 0.0, 0.8}
\definecolor{amethyst}{rgb}{0.8, 0.0, 0.0}
\definecolor{green}{rgb}{0.1, 0.8, 0.}

\usepackage{xcolor} 
\definecolor{darkred}{rgb}{0.6, 0, 0}  

\makeatletter
\renewcommand\normalsize{\@setfontsize\normalsize{12}{14}}
\renewcommand\small{\@setfontsize\small{10}{12}}
\renewcommand\scriptsize{\@setfontsize\scriptsize{7}{8}}
\renewcommand\tiny{\@setfontsize\tiny{5}{6}}
\renewcommand\large{\@setfontsize\large{13}{15}}
\renewcommand\Large{\@setfontsize\Large{14}{16}}
\renewcommand{\thesubsection}{\thesection.\arabic{subsection}}
\renewcommand{\thesubsubsection}{\arabic{subsubsection}}
\renewcommand{\theparagraph}{\alph{paragraph}}
\renewcommand{\thesubparagraph}{\roman{subparagraph}}
\titleformat{\section}[block]{\normalfont\normalsize\scshape\center}{\thesection.}{1em}{}
\titleformat{\subsection}[block]{\normalfont\itshape\center}{\thesubsection.}{1em}{}
\titleformat{\subsubsection}{\normalfont}{\thesubsubsection.}{1em}{}
\titleformat{\paragraph}{\normalfont}{\theparagraph.}{1em}{}
\titleformat{\subparagraph}{\normalfont\small}{\thesubparagraph.}{1em}{}
\titlespacing{\subparagraph}{0pt}{12pt}{12pt}

\titleformat{\section}
  {\bfseries\Large}  
  {\thesection}         
  {1em}                 
  {}
\titlespacing*{\section}{0pt}{10pt}{10pt}  

\titleformat{\subsection}
  {\bfseries}  
  {\thesection.\arabic{subsection}}         
  {1em}                 
  {}
\titlespacing*{\section}{0pt}{20pt}{10pt}  
\AtBeginEnvironment{abstract}{\begin{adjustwidth}{0.cm}{0.cm}}
\AtEndEnvironment{abstract}{\end{adjustwidth}}

\makeatother

\cftsetindents{section}{3em}{3em}
\cftsetindents{subsection}{4em}{3em}
\cftsetindents{subsubsection}{5em}{3em}
\cftsetindents{paragraph}{6em}{3em}
\cftsetindents{subparagraph}{7em}{3em}

\fancypagestyle{mypagestyle}{%
  \fancyhf{}
  \fancyhead[OC]{\textsc{Dust in Galaxy Evolution}}
  \fancyhead[OL]{}
  \fancyhead[OR]{\thepage}
  \fancyhead[EC]{\textsc{Dust in Galaxy Evolution}}
  \fancyhead[EL]{\thepage}
  \fancyhead[ER]{}
  \fancyfoot[O]{}
  \fancyfoot[E]{}
}

\fancypagestyle{mypagestyle}{%
  \fancyhf{}

  \fancyhead[OC]{\textsc{Dust in Galaxy Evolution}} 
  \fancyhead[OL]{}  
  \fancyhead[OR]{}  
  \fancyhead[EC]{\textsc{Dust in Galaxy Evolution}} 
  \fancyhead[EL]{}  
  \fancyhead[ER]{}  

  \fancyfoot[C]{\thepage}

}
\title{\LARGE\textbf{\textsc{Modeling Dust in Galaxy Evolution Simulations}}}

\author{\Large Massimiliano Parente}
\affil{\small SISSA, Via Bonomea 265, I-34136 Trieste, Italy}
\affil{\small INAF, Osservatorio Astronomico di Trieste, via Tiepolo 11, I-34131, Trieste, Italy}
\affil{\small IFPU, Institute for Fundamental Physics of the Universe, Via Beirut 2, 34014 Trieste, Italy}
\affil{\small \href{mailto:mparente270@gmail.com}{\textcolor{blue}{mparente270@gmail.com}} \quad \href{mailto:massimiliano.parente@inaf.it}{\textcolor{blue}{massimiliano.parente@inaf.it}}}

\date{}

\begin{document}

\maketitle
\urlstyle{same}

\vspace{-1em}

\begin{abstract}
\normalsize{
Dust grains play a fundamental role in galaxies, influencing both their evolution and observability. As a result, incorporating dust physics into galaxy evolution simulations is essential. This is a challenging task due to the finite resolution of such simulations and the uncertainties on dust grain formation in stellar envelopes and their evolution in the Interstellar Medium (ISM). This report reviews some of the most commonly used techniques for modeling dust in galaxy evolution simulations, with a particular emphasis on developments from the past $\sim$10 years in both hydrodynamic and semi-analytic approaches. Key findings from these simulations are presented, discussed, and compared to the most recent available observations. These include the dust-to-gas vs metallicity relation, the abundance of dust within and outside galaxies, and predictions on the relative importance of various processes affecting dust. The analysis presented here highlights significant achievements as well as the limitations in our current modeling of dust evolution in galaxies.
}

\vspace{1em}
\begingroup
\renewcommand{\thefootnote}{}  
\footnotetext{This analysis was originally performed as part of my PhD thesis, and this report has been adapted and updated from that work. The thesis -- titled \textit{Dust in hydrodynamic and semi-analytic galaxy evolution simulations} -- is publicly available at \url{https://iris.sissa.it/handle/20.500.11767/141790}.}
\endgroup
\end{abstract}

\vspace{1em}
\section*{\textsc{Contents}}

\begin{itemize}[label={}, left=0pt, labelsep=10pt, itemsep=0pt, topsep=0pt]
\setstretch{0.6}
\small{
  \item 1 Introduction \dotfill \pageref{sec:intro}
  \item 2 Dust life-cycle \dotfill \pageref{sec:dustlifecycle}
  \item 3 Chemical evolution models \dotfill \pageref{sec:dust:chemevo}
  \item 4 Modeling dust in numerical simulations \dotfill \pageref{sec:dustmod:numsimu}
  \item 5 Key results from simulations \dotfill \pageref{sec:dust:keyresult}
  \item 6 Future prospects for dusty simulations \dotfill \pageref{sec:futureprosp}
}
\end{itemize}

\section{Introduction}
\label{sec:intro}

Dust plays a crucial role among the various components of the Universe, especially in the context of galactic and extra-galactic astrophysics. Dust grains are small solid particles ranging in size from $ \sim$ \r{A} to $ \sim 0.1\, \mu {\rm m}$, composed of heavy elements such as carbon, oxygen, magnesium, silicon, and iron. Despite constituting a tiny fraction of the galaxies Interstellar Medium (ISM), dust is critically important for both the detectability and evolution of these systems.

As for detectability, grains significantly obscure light emitted by stars and Active Galactic Nuclei (AGN), making it necessary to consider dust when interpreting observations. This insight dates back a century ago to \citet{Trumpler1930} study on light absorption in our galaxy, which linked it to the presence of ``fine cosmic dust particles of various sizes''. The author also noted the non-uniform distribution of these particles within the galaxy.
Dust grains absorb radiation in the optical-UV spectrum, subsequently re-emitting it in the infrared (IR) region, contributing a substantial fraction to the total panchromatic emission of the galaxy.

Detecting this IR emission became a primary objective for numerous facilities developed since the 1980s, such as the Infrared Astronomical Satellite (IRAS) and the Infrared Space Observatory (ISO), which operated within the wavelength range of $\sim 10 \, {\rm to} \, 250 \, {\mu \rm m}$. These instruments confirmed the great potential of observing the sky in the IR wavelength regime, where dust-obscured galaxies or regions become visible.
In more recent times, we have a multitude of instruments able to detect dust emission across (observed) IR and longer wavelengths (sub-millimeter, radio), like the \textit{Herschel} Space Observatory and the Atacama Large Millimeter/submillimeter Array (ALMA), extending our ability to observe galaxies even at high redshifts. These advancements, coupled with  multi-wavelength imaging from other facilities, have significantly enhanced our understanding of galaxy evolution and the role of dust in shaping their observable properties. 

However, the interaction between radiation and dust depends on numerous uncertain factors, including the relative star-dust geometry\footnote{In this context, the term geometry refers to the stellar age-dependent relative spatial distribution of stars and the dusty medium, which affect the total outcoming SED of a galaxy \citep[e.g.][]{Silva1998}.} and the microscopic properties of dust grains, such as their size and chemical composition. Therefore, it is essential to study the nature of astrophysical grains. Many approaches have been taken to link observed radiation with grain properties based on physical models of the interaction between these tiny particles and light. Examples include observing dust in stellar envelopes or ejecta, where dust grains are thought to form \citep[e.g.][]{Matsuura2011}, or examining the UV extinction \citep[e.g.][]{Fitzpatrick2007}. Additionally, understanding the chemical composition of dust is possible by studying the depletion of metals missing from the ISM gas phase \citep[e.g.][]{Jenkins2009} and examining meteorites \citep[e.g.][]{Lewis1987,Bernatowicz1987}. However, such detailed studies on dust are only available in the Milky Way (MW) and, to some extent, in nearby galaxies, making the interpretation of the observations of dusty galaxies at high redshift even more uncertain.

Besides affecting the radiation we observe from galaxies, dust grains play a significant role in the physical processes that shape galaxy evolution. They contribute to the cooling of hot gas: when ions collide with dust grains, thermal energy is transferred to the grains, which then radiate this energy in the infrared, thus easily losing it from the system \citep[e.g.][]{Burke1974,Dwek1981,Montier2004}. On the other hand, metals removed from the gas by dust grains would serve as effective coolants in their gaseous form. Dust grain surfaces also catalyze the formation of H$_2$ molecules, along with other reactions \citep[see review by][]{Wakelam2017}. Consequently, dust promotes the formation of molecular clouds, which are the sites of star formation, and shields them from stellar light.
The presence of dust in the ISM can enhance the effectiveness of radiation pressure by orders of magnitude, stimulating the onset of galactic winds \cite[e.g.][]{Murray2005}.
Dust can also help in creating a reservoir of low angular momentum gas in galaxies, providing accretion fuel for their supermassive black holes \citep[SMBHs, e.g.][]{Granato2004}.\\

In summary, dust is a crucial ingredient of galaxies, and understanding its formation and evolution is necessary for interpreting observations and modeling galaxy evolution.\\

For this reason, it is highly desirable to model its production and evolution within galaxy evolution simulations. 
Although various types of simulations exist today, this report will focus on two approaches: semi-analytic models (SAMs) and hydrodynamic simulations. Both methods aim to describe the properties and evolution of galaxies within a cosmological framework by incorporating physical processes believed to shape galaxy formation and evolution. The fundamental difference between these approaches lies in the smallest scales they can resolve: hydrodynamic simulations feature a better resolution but are significantly more computationally expensive.\\
Processes occurring below a given resolution (in mass or spatial scale) are typically included using sub-grid modeling, a widely used strategy in numerical simulations (for a comprehensive review, see \citealt{Somerville2015a}). The modeling of dust is one such process that requires a sub-grid treatment. In this context, a dust model is a physical model designed to predict the abundance and properties of dust throughout the galaxy evolution simulation. This model is fully integrated into the simulation, meaning it both is based on and influences the simulation predictions.

While most galaxy evolution simulations do not explicitly model the formation and evolution of dust grains,\footnote{In many cases, when dust predictions are required, a constant dust-to-metal ratio is assumed, allowing dust mass estimates to be derived from the chemical predictions of these simulations.} An increasing number of studies have started incorporating such treatments. Three key motivations drive such a growing interest by the community. First, explicit modelling of dust helps us understand the role of various processes that influence the abundance of observed dust in both local and high-redshift galaxies. Second, it is a necessary step for our simulations to properly predict observable quantities, which are of course affected by both dust obscuration and emission. Lastly, since dust grains actively participate to the physics of galaxy evolution, modeling dust is essential to incorporate a number of processes that improve the accuracy of our simulations.\\

The aim of this report is to examine the development, evolution and predictions of dust models applied to galaxy evolution simulations. The starting point of any dust model lies in the assumed life cycle of dust grains in the galactic environment, which is sketched in Section \ref{sec:dustlifecycle}. Section \ref{sec:dust:chemevo} reviews early dust models used in galactic chemical evolution studies. Section \ref{sec:dustmod:numsimu} introduces and compares dust models currently employed in both SAMs and hydrodynamic simulations, emphasizing the similarities and differences among different groups. In Section \ref{sec:dust:keyresult}, I will present comparisons between simulations in terms of key predictions, such as dust abundance, the relationship between dust and metallicity, and the role of various processes included in the simulations. Additionally, I will discuss potential discrepancies between different models and between models and observational data. Finally, Section \ref{sec:futureprosp} highlights possible future directions for dust modeling in galaxy evolution simulations.

\section{Dust life-cycle}

\label{sec:dustlifecycle}

\begin{landscape} 
    \begin{figure}
    \centering
    \includegraphics[width=0.75\columnwidth]{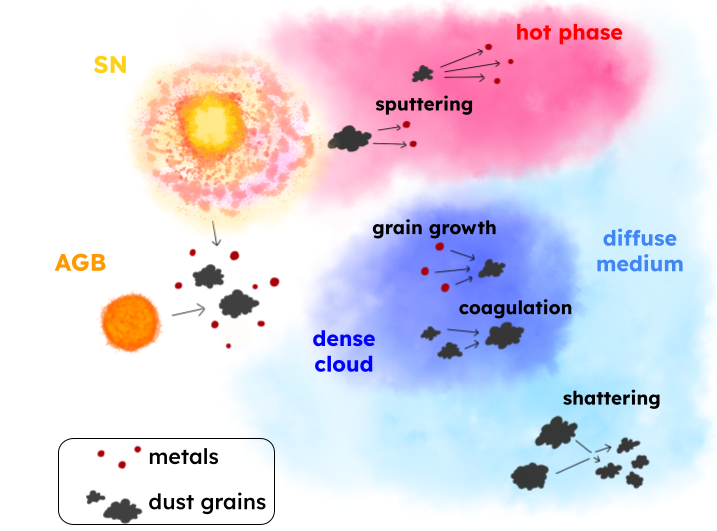}
    \caption[Schematic sketch depicting the dust life-cycle in galaxies]{Schematic sketch depicting the dust life-cycle in galaxies. Grains form in AGB winds and SNe ejecta, and are thus injected into the ISM. Here, accordingly to the physical conditions of the medium, different processes may affect grains evolution. Shattering in the diffuse and turbulent medium, coagulation of small grains and accretion of gas phase metals in dense and cold clouds, sputtering in the hot medium. See text for details on the processes.}
    \label{fig:lifecycle}
    \end{figure}
\end{landscape}

To model the evolution of dust grains within the broader galactic context, it is essential to understand their life-cycle and how various astrophysical processes influence their abundance and properties. A schematic illustration of the main processes described below is provided in Figure \ref{fig:lifecycle}.

Dust grains are produced in the final stages of stellar evolution. As suggested by IR observations of circumstellar material \citep[e.g.][]{Gehrz1971}, high density winds associated with late stellar phases, in particular the AGB one, of low-to-intermediate mass stars ($1 \lesssim M/M_\odot \lesssim 8$) are suitable sites for the condensation of solid grains. The composition of these grains depends on the chemical composition of the outflowing material \citep[e.g.][]{Ferrarotti2006,Gail2009,DiCriscienzo2013,Nanni2013,Nanni2014,Dell’Agli2017}{}{}. Grains produced in AGB winds are expected to be relatively large, with sizes $\gtrsim 0.1\,{\mu \rm m}$ \citep[e.g.][]{Winters97,Ventura2012}. 

The other relevant channel of grains production are SNe ejecta. Far-IR observations of the SN1987A event have clearly demonstrated this \citep{Moseley1989, Kozasa1989,Wooden1993}. Models based on the nucleation theory\footnote{The nucleation theory describes the formation of dust grains in supernova outflows through the condensation of gas-phase atoms and molecules into solid clusters. This process depends on temperature, density, and supersaturation conditions, governing the initial dust formation and subsequent grain growth.} \citep{Feder1966} provided estimates of the amount and properties of grains formed in ejecta of core collapse SNe, predicting large amount of dust formed ($\simeq 0.1-1\,M_\odot$ per event; e.g. \citealt{Todini2001,Nozawa2003}). However, the survival of these grains during the explosion-driven reverse shock is uncertain \citep[e.g.][]{Nozawa2007,Bianchi2007,Bocchio2014}{}{}. The reverse shock is also expected to skew the distribution of grains towards large sizes, since small grains are destroyed more easily \citep[e.g.][]{Nozawa2007}{}{}.\\

\newgeometry{margin=2cm} 

\restoregeometry

Once ejected into the ISM, various processes can influence the evolution of dust grains. Collisions between grains can affect the grain size distribution. In dense and cold molecular clouds, low-speed collisions promote the sticking of grains, known as \textit{coagulation}, resulting in the formation of larger particles \citep[e.g.][]{Ormel2009,Hirashita2014}{}{}.  In contrast, in the diffuse and turbulent medium, collisions can lead to the fragmentation, or \textit{shattering}, of grains \citep[e.g.][]{Yan2004, Hirashita2009}{}{}, which significantly shift the distribution towards smaller sizes.

Dust grains can also change their mass, shape and chemical composition during their journey into the ISM. In dense and cold clouds, gas-phase metals can stick on the surfaces on dust grains, increasing their mass \citep[e.g.][]{Liffman1989,Dwek1998,Hirashita1999}{}{}. This process is often referred to as \textit{grain growth} or \textit{(metals) accretion}. It is of course favoured by the metals content of the cloud, and it also depends on the grain size distribution. Indeed, this is a surface process and it is more likely to occur on small dust grains, whose surface-to-volume ratio is larger \citep[e.g.][]{Hirashita2011}{}{}.

In hostile environments, energetic collisions between gas and grains can cause grains to lose mass by returning gas-phase metals to the ISM, a process known as \textit{sputtering}. The most destructive environment for this process is the shock generated by a SN explosion, where conditions for both thermal and non-thermal sputtering are met, such as high velocities $\gtrsim 100\,{\rm km\,s^{-1}}$ and high temperatures $\gtrsim 10^6\,{\rm K}$ \citep[e.g.][]{Dwek1980,McKee1989,Nozawa2006,Hu2019}. In this work, this process will be referred to as \textit{SN destruction} to differentiate it from \textit{pure} thermal sputtering, which occurs in the hot gaseous galactic and circum-galactic phases. This occurs in the high-temperature ($T\gtrsim10^6\,{\rm K}$) plasma as a result of ions-grains collisions which eventually leads to grains erosion \citep[e.g.][]{Barlow1978,Tsai1995}{}{}. 
Another (minor) destruction process at work for dust grains is photo-evaporation, which occurs in the presence of strong radiation fields \citep[e.g.][]{Nanni2024}{}{}.\\

To conclude, it is important to mention some additional properties of grains that influence their life-cycle in the ISM. Grains typically possess a non-zero electric charge, which can affect their interactions with ions (and thus grain growth), their dynamics, and their interactions with photons \citep[e.g.][]{Zhukovska2018, Melzer2019, Glatzle2022}. Another characteristic that can impact grain evolution in the ISM is porosity, as it alters the total surface area \citep[e.g.][]{Hirashita2022}. In this report, as well as in the large majority of galaxy evolution simulations which include dust, neither electric charge nor porosity are considered.


\section{Chemical evolution models}

\label{sec:dust:chemevo}

Historically, the first tools adopted by theoreticians to model dust formation and evolution in a galactic context were chemical evolution models (e.g. \citealt{Tinsley1980}; see also \citealt{Matteucci03} and references therein). These are typically one-zone and one-phase models. The former means that they lack spatial resolution (i.e. same properties in the whole disc, or disc+bulge in the case of two-zones models), while the latter that the model averages gas ISM properties over the cold, warm and hot phases. In the most simple cases, a set of coupled non-linear integro-differential equations are used to model gas infall, star formation, and chemical enrichment, eventually adopting different prescriptions for different types of galaxies \citep[e.g.][]{Matteucci87,Bradamante98,Chiappini2001}{}{}.\\

Due to the chemically-oriented nature of these models, their use for studying galactic dust was a natural choice. The first pioneering attempts to include dust within basic models can be traced back to the 1980s \citep{Dwek1980, Liffman1989, Liffman1990, Wang1991}. These early studies aimed mainly to explain the depletion patterns or the chemical composition of grains in the solar neighbourhood. They provided valuable insights for future, more complex models. For example, they suggested the importance of SNe as grains factories\footnote{In \cite{Dwek1980}, several sources were assumed to produce grains: planetary nebulae, winds of red giant stars, proto-stellar nebulae, novae and supernovae. As they reported in a note added as proof, there was an observational claim of dust condensation in SN ejecta. The future clear SN1987a detection was just the confirm of the relevance of the SN channel.}, as well as the need of grain growth in molecular clouds, which was not always included at that time.

\citet{Dwek1998} was the first to include a dust model in a proper chemo-dynamical, two-zones chemical evolution model. The equation describing dust evolution is of the type\footnote{Actually, in the original \citet{Dwek1998} model dust surface densities are considered instead of dust masses.}:

\begin{equation}
    \dot{M}_{\rm dust} = \dot{M}^{\rm stars}_{\rm dust}(t) - \frac{M_{\rm dust}}{M_{\rm gas}}{\psi} + \dot{M}^{\rm growth}_{\rm dust}(\tau_{\rm gg}) - \dot{M}^{\rm SN \,des}_{\rm dust}(\tau_{\rm SN\,des}) - \dot{M}^{\rm outflow}_{\rm dust}.
\end{equation}

This equation relates the ISM dust mass variation to the time-dependent stellar production of grains $\dot{M}^{\rm stars}_{\rm dust}(t)$, the mass of dust re-incorporated into stars (commonly referred to as \textit{astration}) proportional to the SFR $\psi$, the growth of grains in molecular clouds ($\dot{M}^{\rm growth}_{\rm dust}$, dependent on some timescale $\tau_{\rm gg}$), the destruction of grains in SN shocks ($\dot{M}^{\rm SN\,des}_{\rm dust}$, dependent on some timescale $\tau_{\rm SN\,des}$), and the mass of dust removed by outflows $\dot{M}^{\rm outflow}_{\rm dust}$. These processes are identified as the fundamental ones in all the future dust models which aim at following the amount of dust\footnote{Models that additionally describe the evolution of grain sizes need to take into account other processes, such as shattering and coagulation.}, even in the case of more complex, numerical simulations. The main differences among various dust models stand in the assumptions (e.g. dust yields, time-scales functional forms) that determine the terms in the above equation, along with possible adjustments in multi-phase models.\\

The seminal work by \citet{Dwek1998} focused on galactic dust, particularly on depletion patterns. The model confirmed the dominance of SNII (and SNIa, for iron dust) as dust sources and grain growth in molecular clouds.\\
Thanks to the advance in observations, particularly the \textit{IRAS} satellite and SCUBA camera operating in the IR, it also became possible to study dust beyond our Galaxy. A particularly intriguing relation that emerged was that between the dust-to-gas ratio (DTG) and gas phase metallicity $Z$ \citep[e.g.][]{Issa90,Schmidt1993}{}{}. Chemical models constituted a powerful tool to investigate this relation in both dwarf and spiral local galaxies \citep[][]{Lisenfeld1998,Dwek1998,Hirashita99,Hirashita1999, Edmunds2001,Inoue2003,Galliano2008,Mattsson2012,Kuo2013,Feldmann2015,Schneider2016}{}{}. 
In the low metallicity range, dwarfs exhibit a flatter and more scattered pattern in the observed DTG$-Z$ relationship, whereas spiral galaxies display an inclining slope. It became apparent that, while the relation in dwarf galaxies is well reproduced thanks to the stellar production channel and some stochasticity to ensure the observed scatter, the growth in clouds is necessary for explaining the DTG in more metal enriched galaxies \citep{Hirashita99,Hirashita1999}. \citet{Inoue2011} and \citet{Asano2013} introduced the concept of \textit{critical metallicity}, above which ISM grain growth becomes the dominant process. This is set by the competition between the efficiency of grain growth and destruction (or dust-free inflows, see \citealt{Feldmann2015}).\\

Other extra-galactic observations \citep[e.g.][]{Lilly1999,Eales2000}{}{} made it necessary to expand the study of dust evolution in galaxies different from our MW. A step in this direction was moved by \citet{Calura2008}. Adopting the formalism of \citet{Dwek1998}, they formulated a treatment of the evolution of dust in spirals, ellipticals and irregular galaxies, assuming different prescriptions for grain growth in each of these categories. Also, they included the ejection of dust by outlaws, pointing out that large dust masses may be found in the hot CGM/IGM, although they were missing a model for the destruction in this hostile medium by thermal sputtering.

Building on the \citet{Calura2008} model, \citet{Gioannini2017} tried to overcome the lack of cosmological context of chemical models. The different treatment of dust evolution in different galaxy types are coupled with some assumptions on the abundance of spiral, elliptical and irregular galaxies across cosmic time. In this way, the authors could predict the cosmic dust mass density, $\Omega_{\rm dust}$, which was in good agreement with the observations available at that time. This is the first time a model has been compared with this quantity, which will be widely discussed by the forthcoming cosmological models with dust in the subsequent years.\\

One of the key components of these models is the stellar production of grains, which is also very uncertain. The first detailed studies, based on the nucleation theory, provided predictions on the effective grains production by different stellar sources (AGBs: \citealt{Ferrarotti2006}; and SNe \citealt{Todini2001,Nozawa2006,Bianchi2007}). \citet{Zhukovska2008} were the first to implement detailed mass and metallicity-dependent dust yields from AGB stars in a chemical evolution model. 

The focus on dust sources was further motivated by high-z observations, thanks to which a picture emerged in which a high abundance of dust was already in the first $\lesssim 1\,{\rm Gyr}$ of the Universe lifetime. This is the case of the SDSS1048+46 quasar at $z=6.2$ \citep{Maiolino2004}, which challenged the models of the time \citep{Maiolino2004a,Dwek2007}, who supported a picture in which SN production has to dominate in these early epochs, although with some extreme assumptions. The model by \citet{Valiante2009}, based on detailed dust yields, suggested that stellar sources could account for the observed dust, and also that the AGB contribution is not negligible at that early times. However, in their model they neglect grains accretion in MCs. Studying $z\gtrsim 6$ FIR-detected QSOs, \citet{Calura2014} claimed the need for a strong efficiency of grain growth (or a strong star formation efficiency or a top-heavy IMF; see also \citealt{Rowlands2014}). A similar conclusion was reached by \citet{Valiante2011} and \citet{Mancini2015}, with a semi-numerical\footnote{Here semi-numerical means that the chemical model has been applied to a set of merger trees obtained with the extended Press-Schechter algorithm.} chemical model.\\

Another important aspect often missing in many chemical evolution models was the treatment of the grain size distribution. Some seminal works studied the variations of the latter in different processes, such as sputtering, grains growth, shattering and coagulation \citep[e.g.][]{Liffman1989,O’Donnell1997,Hirashita2010,Hirashita2010a}{}{}. However, the first work including a treatment of the processes affecting dust sizes in a galactic context, is the one by \citet{Asano2013a}. Stars produce large ($\sim 0.1\,{\mu \rm m}$) grains, whose size may evolve by shattering and coagulation, modelled in a simplified way. Also, ISM processes like growth and destruction are size-dependent. The result is that the grain size distribution evolves with the galaxy age, from a large-grains to a small-grains dominated one. Moreover, dust abundance is also affected by the processes: shattering plays a relevant role in producing small grains and boosting grain growth, which is more effective on small-sized particles. 

Cheaper treatments of grain size distributions introduced in the context of chemical models are also the two-size approximation \citep{Hirashita2015} and the moments method\footnote{The method of moments is a statistical technique that involves tracking (some of the) moments of the mass and size distribution of grains. This approach approximates and reconstructs the size distribution from a truncated hierarchy of moments.} \citep{Mattsson2016}.\\

Recently, despite the advent of complex models that have diminished the role of chemical models, they are still used to study the dust budget and scaling relations of galaxies \citep[e.g.][]{Clark2015,Ginolfi2017,DeVis2017,Vis2019,Nanni2020,Galliano2021a,Calura2023}{}{}. Also, one-zone dust-focused\footnote{Their goal is to study the dust evolution of galaxies based on certain assumptions about key features, such as the star formation history.} models are adopted to explore dust-related processes still not included in galaxy evolution models. Examples are dust grains aromatization \citep{Hirashita2020a}, the treatment and impact of grain porosity \citep{Hirashita2022a}, the evolution of small carbonaceous grains \citep{Hirashita2022}, molecules formation on grains \citep{Hirashita2023}.\\

In summary, chemical evolution models provided a valuable tool for the pioneering exploration of dust evolution in galactic contexts. They obtained key results, such as a physical explanation of the DTG$-Z$ relation and the evolution of the grain size distribution. Moreover, these models developed analytical prescriptions for dust-related processes which served as foundation of the forthcoming numerical simulations.

\section{Modeling dust in numerical simulations}

\label{sec:dustmod:numsimu}

Both SAMs and hydrodynamic simulations allow us to study dust evolution in a cosmological context, and sometimes on (sub-)galactic scales. The importance of these tools has to be understood in light of the large amount of observations now available. Nowadays, instruments and facilities (e.g. \textit{Herschel}, ALMA) make it possible to make a detailed census of dusty galaxies up to very high redshifts and study dust distribution in resolved nearby galaxies. Numerical simulations allowed our community to compare with these data and to test our dust models against them.\\

As most of the physics considered, given the limited resolution of hydrodynamic simulations and even more of SAMs, dust-related processes must be implemented in a \textit{sub-grid} fashion, relying on approximated relations. Drawing from more than 20 years of experience with chemical evolution models has been fundamental in developing these recipes, which, however, required some adaptation to fit the new numerical frameworks. The following section is devoted to detail in this practical aspect.\\
The dust model recipes are tailored to align with the basic working scheme of SAMs and hydrodynamic simulations, which define the fundamental resolution units of these frameworks. I will resume the main concepts adopted by most of the simulations to regulate the exchange and the evolution of dust grains in such \textit{units}. A comprehensive summary of the processes modeled in various published simulations is reported in Table \ref{tab:subgriddust}.

\subsection{Stellar Production}

The starting point of dust life-cycle is the stellar production of grains, and their subsequent ejection in the gaseous medium. This process closely resembles those of metals production (by stellar sources) and spreading. To treat it, a fraction of the dust composing ejected metals are labelled as \textit{condensed} in the solid phase, that is, they are in dust grains. These will enrich the gaseous component of a semi-analytic galaxy (e.g. the gaseous disc, the hot halo) or the gas close to stars in the hydrodynamic simulation. In most cases, numerical simulations are provided with detailed chemical models which can associate to each stellar population a specific SFH, thus they can follow the delayed chemical (and dust) enrichment by different stellar sources (e.g. SNII, AGBs, SNIa).\\

Therefore, the fraction of metals \textit{condensed} into dust grains is the first important assumption for any dust model. Inspired by chemical evolution models, different options are available. In the most simple cases, it is assumed that a fixed fraction $\delta$ of all metals (without tracking each chemical element individually) is ejected in the solid phase \citep[e.g.][for SNII only]{Aoyama2017}. A more sophisticated but still simple approach is that of \citet{Dwek1998}, widely adopted by both SAMs and hydrodynamical simulations \citep[e.g.][]{Bekki2013,Popping2017,McKinnon2016}. In this case, the condensed fraction $\delta^j_i$ (in the broad range $\sim 0.05-1$) is different for each stellar source (\textit{i}; AGBs, SNII, and SNIa) and it refers to each single element assumed to be part of dust grains (\textit{j}; typically C, O, Mg, Si, Fe, but also Al, S, Ca, Ti). 
Another solution is to adopt a condensation fraction which depends on the assumed chemical composition of dust grains. This method, often referred to as \textit{key element} and first introduced by \citet{Zhukovska2008}, consists of identifying, for each dust compound, the element whose abundance is such that it constitutes a bottleneck for the formation of the compound with given stoichiometric ratios. In other words, the key element is the one for which the ratio between the available number of atoms and the number of atoms in a given compound $N/N_{\rm comp}$ is minimum.

A minority of simulations \citep[e.g.][]{McKinnon2018,Graziani2020}{}{} uses instead a treatment of dust production in principle more shophisticated by adopting mass- and metallicity-dependent yields for AGB stars and SNe. However, these yields are based on several assumptions, such as the micro-physics and the thermodynamics of the ejected material, which makes model prediction uncertain.

\subsection{Dust spreading in the ISM}

Newly produced dust grains are subsequently spread in the gaseous medium. However, the dynamics of dust grains is not necessarily the same of the gas.

All the SAMs discussed here assume that dust is simply a component of the various gaseous units (e.g. disc, halo). It is then implicitly assumed that dust is perfectly coupled with gas. This is somehow forced to be in SAMs, due to their lack of spatial resolution. However, it is noteworthy that certain recent SAMs which incorporate dust \citep[][]{Parente2023,Yates2023}{}{} feature coarse spatial resolution of galaxy discs and a simplified treatment of gaseous inflows resulting from angular momentum loss. Therefore, there is an opportunity to introduce a basic model for gas-dust drag interactions. Employing a distinct inflow mechanism for dust could provide a straightforward way of representing this phenomenon.

As well, the large majority of the hydrodynamic simulations mentioned here include dust as an ingredient, or a property, of gas particles. Some exceptions are the work by \citet{Bekki2015,McKinnon2018,Li2021}, in which dust grains are represented by independent \textit{live} particles\footnote{The expression \textit{live particles} has been introduced by \citet{Bekki2015}. In these cases, the gas properties (e.g. thermodynamical, metallicity, SN rates; see later) needed to evaluate the efficiency of dust processes, are evaluated by smoothing the properties of the neighboring gas particles of each \textit{live} dust particle.}. These are coupled to the gas via the modelling of a gas-dust drag force which depends on the gas properties. This method is particularly suitable for studying dynamical dust processes (e.g. radiation pressure; see \citealt{Bekki2015}). Also, these experiments demonstrated that the widely adopted strong coupling between gas and dust is a reasonable assumption, at least at the resolution achievable in cosmological simulations \citep{McKinnon2018}.\\

\subsection{Grains accretion in dense medium}
Some of the aforementioned chemical models generally inspire the treatment of dust grains accretion in simulations \citep[e.g.][]{Dwek1998,Zhukovska2008,Asano2013}, although the practical implementation greatly vary. A commonly adopted functional form for the accretion time-scale is

\begin{equation}
    \tau_{\rm acc} \propto n^{-1} T^{-1/2} Z^{-1},
\end{equation}

where $n$, $T$, and $Z$ are the density, temperature and metallicity of the gas. This formulation takes into account that gaseous medium with higher densities, lower temperatures and higher metal abundances favour the collision and subsequent sticking between gas-phase metals and grains. This process is of course allowed to occur in the cold gas which is expected to host dense medium. However, it is worth reminding that this dense medium, namely molecular clouds, is far below the resolution of our current cosmological simulations. As a result, different approaches are adopted to compute the above quantities. In the semi-analytic approach, only the cold disc is considered, and the density and temperature of the gas are assumed to be fixed \citep[e.g.][]{Popping2017}. In the hydrodynamic approach, these quantities may be evaluated on a particle-by-particle basis, relying on the predictions of the hydrodynamic solver \citep[e.g.][]{McKinnon2016,Li2019,Graziani2020}. In some other cases, being molecular dense clouds below the resolution of typical cosmological simulations, a fixed $n$ and $T$ is assumed \citep[e.g.][]{Granato2021}. \\
The dependence on the dense medium is sometimes explicitly considered, for example re-scaling the timescale by the molecular fraction in the cold gas. This is assumed to be fixed (e.g. \citealt{Aoyama2017}) or computed by mean of sub-grid recipes \citep[e.g.][]{Granato2021,Parente2023}. Alternatively, \citet{Vijayan2019} and \citet{Yates2023} model an exchange of dust between dense and diffuse medium \citep{Zhukovska2014}. \\
In both the semi-analytic and hydrodynamic approach, the metallicity of the gas $Z$ is self-consistently computed by the simulation. However, some other dependences are sometimes adopted (e.g. the mass of dust in molecular clouds \citealt{Vijayan2019}; \citealt{Yates2023}).\\

Eventually, some models incorporate a more sophisticated approach which consider some extra-dependence on grains properties \citep[e.g.][]{McKinnon2018, Granato2021, Li2021}. For example, accretion depends on both the chemical composition (which regulates the mass density) and the size of grains. When dust properties are not considered \citep[e.g.][]{Li2019,Triani2020}, representative values are chosen (e.g. a size of $\sim 0.01 \, \mu{\rm m}$ and material density $\sim 2-3 \, {\rm g\, cm^{-3}}$). When the simulation follows the chemical composition and size of grains, appropriate values are used to compute $\tau_{\rm acc}$. The size of the grains significantly influences the process, with smaller grains being more efficiently accreted ($\tau_{\rm acc}\propto {\rm size}$, e.g. \citealt{Hirashita2011}).\\

\subsection{Grains destruction in SN shocks}
There are relatively few variations in the modeling of dust grain destruction in SN shocks in different simulations. Although the practical implementation may differ, most models are based on the work of \citet{McKee1989}, according to which grains within the sweeping radius of an SN explosion are destroyed. This is conveniently quantified as a swept mass of gas $M_{\rm sw}$, which depends on the energy of the SN explosion and the shock velocity, assumed to be fixed values (of the order of $\sim 10^{51}\, {\rm erg}\,{\rm and}\, \sim 200 {\rm km \, s^{-1}}$). In both SAMs and hydrodynamic simulations, the resulting destruction timescale reads:

\begin{equation}
    \tau_{\rm des, SN} \propto \frac{1}{R_{\rm SN} M_{\rm sw} \epsilon},
\end{equation}

where $R_{\rm SN}$ is the SN rate which is consistently derived from the simulation, and $\epsilon$ a parameter quantifying the dust destruction efficiency of the order of $\sim10\,\%$ \citep[e.g.][]{McKee1989,Nozawa2006}. In some cases, different kind of SNe are associated with different destruction efficiencies \citep{Graziani2020} .\\

The impact of SN explosions on the surrounding medium also depends on the properties of that medium. As demonstrated by \citet[][see also \citealt{Nozawa2006}]{Yamasawa2011}, the swept mass decreases with increasing metallicity and density of the surrounding medium, as more efficient cooling leads to easier shock deceleration. This has led various simulators to adopt the fitting formula provided by \citet{Yamasawa2011} to account for the density (in hydrodynamic simulations) and metallicity dependence of the SN destruction efficiency ($M_{\rm sw} \propto n^{-0.202}Z^{-0.298}$; e.g., \citealt{McKinnon2018}, \citealt{Triani2020}, \citealt{Parente2023}, \citealt{Yates2023}). Furthermore, \citet{Hu2019} noted that silicate grains are more easily destroyed than carbonaceous grains. This has been incorporated into some models that track the chemical evolution of dust \citep[e.g.,][]{Yates2023, Dubois2024}.\\ 

Lastly, models accounting for grain size distribution often take into account the larger efficiency of grains destruction on smaller grains, which may be easily eroded by the intervening shock \citep[e.g.][]{McKinnon2018, Aoyama2020,Dubois2024}{}{}.

 \subsection{Grain size evolution: shattering and coagulation}
Another crucial property of the grains population is its size distribution, as it also affects the rates of various processes. Among the simulations that account for this, two common strategies are used. The first, a coarse and computationally inexpensive method, is the two-size approximation \citep{Hirashita2015}. This approach considers only two representative sizes of grains (e.g., $a \sim 0.005 \, \mu{\rm m}$ and $\sim 0.1 \, \mu{\rm m}$), and it is able to reproduce the main features of full calculations of grain size distribution. The distinction between small and large grains is determined by noting that when considering the full-size distribution, different processes generate two peaks in the distribution, the boundary being located at $\sim 0.03 \, {\mu \rm m}$ (see Figure 6 in \citealt{Asano2013a}). Stars produce large grains, accretion acts on small grains, and eventually, destruction processes have different efficiencies based on the assumed grain size ratio. In more sophisticated approaches \citep[e.g.][]{McKinnon2018,Aoyama2020,Li2021}{}{}, each dust grains population is represented by a $N-$bins size distribution. \\

The processes that regulate the exchange between small and large grains reshuffling the grain size distribution (GSD) without changing the total dust mass, are shattering and coagulation. These are collisional processes where grains, depending on their relative velocity, either fragment into smaller particles (shattering) or stick together to form larger grains (coagulation). The relative velocity of the grains is modelled based on the properties of the gas. Shattering typically occurs in diffuse medium (gas with $n < 1 \, {\rm cm^{-3}}$), while coagulation is effective in dense, unresolved molecular clouds. Therefore, similar to the approach adopted for accretion, coagulation efficiency is often linked to the fraction of dense gas assumed by the model \citep[e.g.,][]{Aoyama2017, Granato2021}. Typical grain velocities are assumed to be approximately $\sim 10 \, {\rm km \, s^{-1}}$ in diffuse gas and $\sim 0.1 \, {\rm km \, s^{-1}}$ in dense gas. However, more complex size- and density-dependent functional forms are used when $N$-bin distributions are considered \citep[e.g.,][]{McKinnon2018, Aoyama2020}. In this context, it is worth highlighting the work of \cite{Narayanan23}, who developed a framework for the formation and evolution of Polycyclic Aromatic Hydrocarbons (PAHs) in idealized galaxy evolution simulations. This builds upon the detailed dust model of \cite{McKinnon2018}. Accounting for the physics ruling PAH abundances requires the inclusion of processes such as aromatization and aliphatization\footnote{Aromatization refers to the de-hydrogenation of carbonaceous dust grains under the influence of UV radiation, resulting in the formation of aromatic bonds and rings. Aliphatization is the inverse process, where H atoms accrete onto grains, and it is especially relevant in dense environments.} 
\citep[see, e.g.,][]{Hirashita2020a}, which are also influenced by the interstellar radiation field.

Lastly, it is worth mentioning the SAM by \citet{Parente2023} as it is the only semi-analytic approach that follows two size distributions\footnote{I also mention the work by \citet{Makiya2022}, who studied the size distribution by post-processing the galaxies catalog of a SAM.}. Given the limitations of the semi-analytic approach, the treatment of shattering and coagulation is simplified, relying on the fraction of molecular gas in the cold gas disc to distinguish between the diffuse and dense components of the ISM.

\subsection{Grains sputtering in the hot medium}

In the hot medium ($T\gtrsim 10^5-10^6\,{\rm K}$), grains are eroded by the collisions with energetic particles. The prescription introduced by \citet{Tsai1995} is by far the most widely adopted to model thermal sputtering in galaxy evolution simulation. It suggests a dependence on the density $n$ and temperature $T$ of the hot gas, as well as on the grain radius $a$:

\begin{equation}
    \tau_{\rm spu} \propto \frac{a}{n} \left[\left(\frac{T_0}{T}\right)^\omega+1\right].
\end{equation}

This is an outcome of theoretical computation (\cite{Tielens1994}, but see also \citealt{Nozawa2006}), which essentially predicts a rapidly ($\omega = 2.5$) increasing sputtering efficiency with temperature $T$ up to $T_0$, temperature above which it is nearly constant. This is trivially implemented in simulations. In SAMs, density and temperature are typically computed assuming a hot, virialized gas in the DM halo, extending up to the virial radius. The hot gas is at the virial temperature and at some average density, dependent on the assumed density profile (e.g., uniform density, $\beta$ model; e.g., \citealt{Yates2023}). In hydrodynamic simulations, sputtering is computed on a particle-by-particle basis, with the hot gas having density and temperature predicted by the hydrodynamic scheme.

\subsection{Astration, inflows, mergers, outflows}

It is worth mentioning some processes that re-locate dust within the gas and stellar phases of the simulations. Astration, the process by which grains are embedded again into newly formed stars, reduces the dust in the ISM. In both the semi-analytic and hydrodynamic approach, this is trivially modelled by decreasing the amount of dust in the star forming gas\footnote{Or according to the neighbouring gas particles in the case of \textit{live} dust particles, \citep[e.g.][]{Li2021}. However, in some cases, astration is not included at all \citep{McKinnon2018}.}, typically assuming a constant DTG. A constant DTG is typically assumed also in the case of inflows, outflows, and mergers in SAMs, in which, during this process, a certain amount of dust is transferred from a reservoir to another (e.g. from a cold disc to a hot halo, in the case of outflows). In hydrodynamic simulations, these processes naturally emerge from the dynamics and transformation of gas particles. Consequently, dust does not deserve a specific treatment. However, it is worth noting that \citet{Li2019} destroy dust in wind gas particles, representing outflows generated by stellar feedback.

\section{Key results from simulations}
\label{sec:dust:keyresult}

Galaxy evolution simulations with dust allowed us to produce theoretical predictions for several different observables and study the various contributions of the dust processes to the total dust budget. In this section, some of the key results obtained by dusty simulations from the last decade are presented. Also, some comparisons are performed and discussed.


\subsection{The DTG-Z relation}

 \begin{figure*}
        \centering
        \begin{subfigure}[b]{0.475\textwidth}
            \centering
            \includegraphics[width=\textwidth]{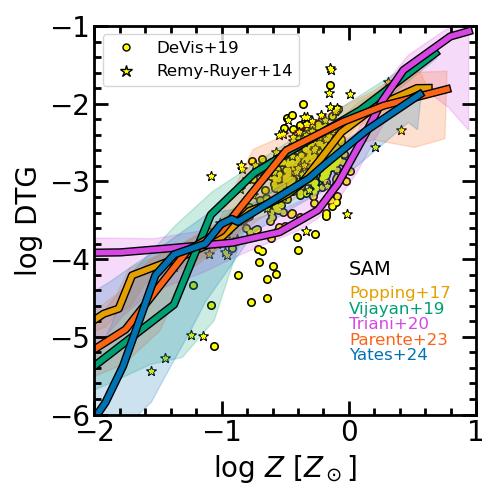}
                
        \end{subfigure}
        \hfill
        \begin{subfigure}[b]{0.475\textwidth}  
            \centering 
            \includegraphics[width=\textwidth]{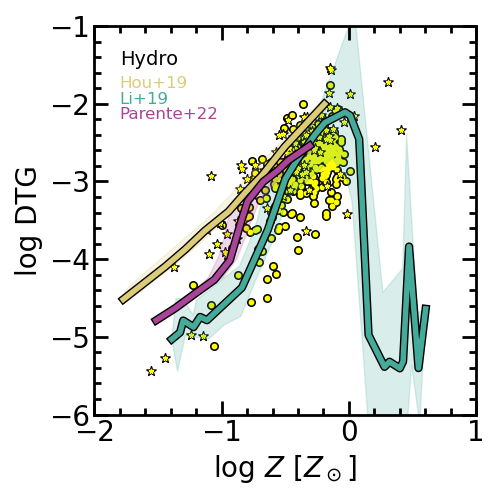}
               
            \label{fig:DTGrev}
        \end{subfigure}
        \vskip\baselineskip
        \begin{subfigure}[b]{0.475\textwidth}
            \centering
            \includegraphics[width=\textwidth]{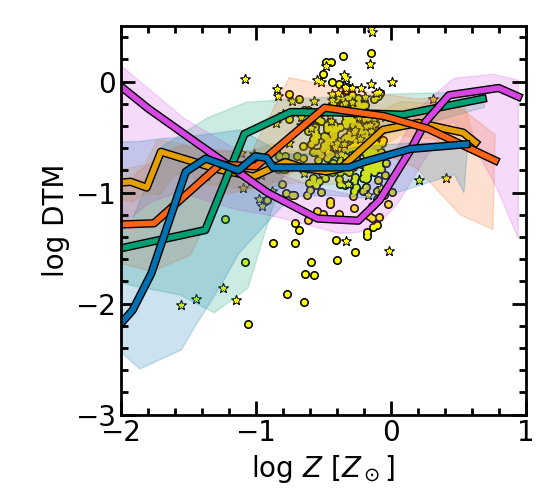}
                
        \end{subfigure}
        \hfill
        \begin{subfigure}[b]{0.475\textwidth}  
            \centering 
            \includegraphics[width=\textwidth]{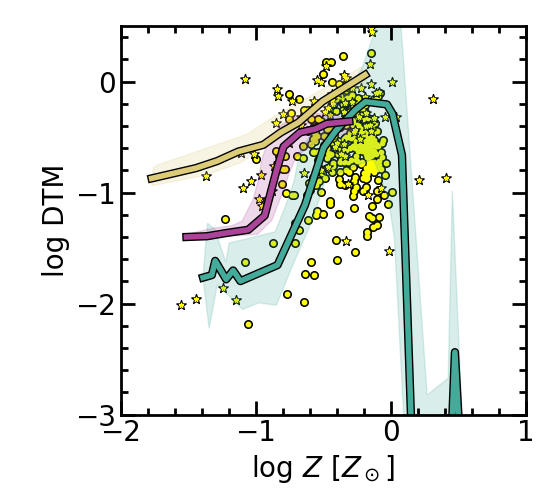}

        \end{subfigure}

        \caption[DTG and DTM versus gas metallicity $Z$ at $z\simeq0$.]{Relation between DTG (top panels) and DTM (bottom panels) versus gas metallicity $Z$ at $z\simeq0$. Predictions from both SAMs (left panels; \cite{Popping2017}, yellow; \cite{Vijayan2019}, green; \cite{Triani2020}, magenta; \cite{Parente2023}, orange; \cite{Yates2023}, their binary stars model, blue) and hydrodynamic simulations (right panels; \cite{Hou2019}, purple; \cite{Li2019}, green; \cite{Parente2022}, dark red) are shown. Median trends and dispersions are reported as solid lines and shaded areas. Observational determinations from \citet{Remy-Ruyer2014} and \citet{Vis2019} are reported in each panel, respectively as filled circles and stars.} 

        \label{fig:DTGrev}
    \end{figure*}

A key diagnostic relation for dust models already discussed in the chemical evolution framework is the DTG (and DTM)-$Z$ relation. Simulations extensively compared with it at both low and high redshift. The result is that models can broadly match the observations up to $z\simeq 5$, but not at all redshifts simultaneously. (\citealt{Popping2022}, but see also \citealt{Yates2023}). Here, I focus on the $z\simeq0$ relation, which is reported in Figure \ref{fig:DTGrev} for various cosmological simulations, compared to observations by \citet{Remy-Ruyer2014} and \citet{Vis2019}.

At low metallicity, DTG and DTM are mainly determined by the stellar production channel and, more specifically, by the condensation efficiency or the adopted stellar yields. This regime has a huge variation ($\sim 2 \,{\rm dex}$) in model predictions. For example, in the study by \citet{Hou2019}, low $Z$ galaxies feature large DTG and DTM ratios, suggesting that their assumed $10\%$ metals condensation fraction in SNII ejecta is too large (or equivalently, the SN-driven destruction is not efficient). A similar conclusion may be drawn for the \citet{Triani2020} model.

Grain growth is the main driver of the peculiar $s-$shape of the increasing relation between DTG (and DTM) and $Z$, \citep[e.g.][]{Popping2017,Li2019,Parente2022}{}{}. This shape highlights the critical metallicity $Z_{\rm crit}$, above which grain growth becomes more efficient than stellar production. The different prescriptions adopted for dust growth, dust production and dust destruction -- as well as the other sub-grid recipes adopted by the models -- may substantially affect the outcomes. Also, in this case, a huge variation is observed in model predictions for $Z_{\rm crit} \sim 10^{-2}-1\,Z_\odot$. \citet{Parente2022} have shown explicitly that the grain growth efficiency regulates $Z_{\rm crit}$. However, \citet{Popping2022} pointed out that different predictions about $Z_{\rm crit}$ may inform about more profound differences among various simulations, such as different star formation timescales \citep{Asano2013,Zhukovska2014,Feldmann2015}. They also suggested that the \citet{Yamasawa2011} $Z-$dependent prescription for dust destruction in SN shocks may be inadequate, since the \citet{Triani2020} model -- which incorporates this prescription -- predicts a (decreasing up to $Z\sim Z_\odot$) DTM-Z relation inconsistent with observations. However, the same $Z-$dependent prescription is adopted in \citet{Parente2023}, which predicts a DTM-Z relation in keeping with observational data. The \citet{Triani2020} inconsistency may result from a too-low grain growth efficiency at relatively high $Z$, which makes the destruction channel dominant. This would agree with the quite high $Z_{\rm crit}\sim Z_\odot$ featured by this model.

At the high $Z$ regime, the saturation of the grain growth is reached. Hence, a $\sim$ constant DTM is retrieved. This is not the case with the \citet{Li2019} model, which predicts a strongly decreasing DTG and DTM in this diagram region. This behaviour is attributed to the AGN feedback, which induces gas heating and consequently enhances dust destruction through thermal sputtering.\\

\subsection{The role of dust-related processes}


\begin{figure}
\centering
\includegraphics[width=0.9\columnwidth]{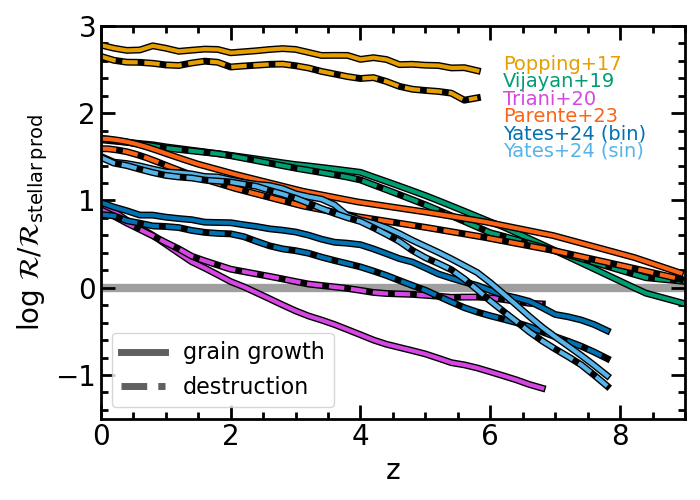}
\caption[Cosmic rate of ISM grain growth and ISM grains destruction.]{Cosmic rate of {ISM} grain growth (solid lines) and ISM grains destruction (dashed lines) as predicted by different SAMs: \citet[][yellow]{Popping2017}, \citet[][green]{Vijayan2019}, \citet[][pink]{Triani2020}, \citet[][orange]{Parente2023}, \citet[][dark and light blue for the model with single and binary stars, respectively]{Yates2023}. Grains destruction refers to only SN destruction in the case of \citet{Triani2020}. The horizontal gray line marks the rates normalization, i.e. the dust stellar production rate ($=1$ in this scale).}
\label{fig:cosmicdustrates}
\end{figure}

To what extent the various processes contribute to the dust budget of galaxies? Predictions from various SAMs over a broad redshift range are shown in Figure \ref{fig:cosmicdustrates}. The grain growth and destruction rate in the ISM is summed over the whole galaxy population and normalized to the dust production rate by stellar sources. This normalization procedure is needed to highlight the relative importance of the ISM processes. All the simulations agree on the increasing dominance of the grain growth channel over the stellar production channel as $z$ decreases. The latter only dominates in the early phases of galaxy evolution, although there is no consensus on the redshift at which this \textit{transition} occurs. This transition redshift is very high ($z \gtrsim 6$) for $5$ out of the $6$ models reported, while \citet{Triani2020} predicts an unusually low one $(z \simeq 2)$. It appears that grain growth is limited in this model\footnote{Note that in their model, a single size of grains is assumed, namely $0.1 \, \mu {\rm m}$. Smaller sizes could enhance grain growth.} (see also discussion of the DTG- and DTM-Z relation).

Also, there is a huge difference in the predictions of the relative importance of grain growth over stellar production, being it $\sim 10$ in \citet{Triani2020} and \citet[][their model with binary stars]{Yates2023} and $\sim 10^3$ in \cite{Popping2017} at $z\sim 0$. Although the adopted grain growth prescriptions are different (see Table \ref{tab:subgriddust}), it is worth noting that these variations might originate from the substantial differences in the underlying galaxy evolution and chemical models. These affect star formation, metallicity, and molecular fraction, which are all relevant in determining the timescales of dust processes. Interestingly, the SAM by \citet{Yates2023}, in which the effect of chemical enrichment by binary stars has been investigated, found two similar \textit{transition} redshifts $(z \simeq 6.5)$ when adopting the chemical yields from both single and binary stars. \\

The clear result that should be emphasized is the relative similarity of the grain growth and destruction rate (but at $z \gtrsim 2$ in \cite{Triani2020}, which supports the hypothesis of limited grain growth in this SAM). This means metal recycling is quite fast: grains destroyed in SN shocks produce gas-phase metals, which are quickly re-locked into grains by grain growth. As a result, large differences in rates ($\sim 2$ order of magnitudes) do not necessarily reflect large differences in dust masses (see discussion later). 

Finally, note that quantities discussed here have been summed over the whole galaxy population. Still, it should be kept in mind that the efficiency of the various processes varies from galaxy to galaxy. For example, \citet{Popping2017} pointed out that grain growth dominates the dust production budget only in relatively massive galaxies (${\rm log} \, M_{\rm stars}/M_\odot \gtrsim 8$).

\subsection{Dust at high$-z$}

The role of various processes is more controversial at high redshift. While the hydrodynamic simulation by \citet{Graziani2020} confirmed the importance of the growth of the grains in the ISM to reproduce the dust abundance of $4<z<8$ ALMA detected galaxies, \citet{Dayal2022} claimed recently a negligible role of this process to explain the mass of dust in REBELS $z \simeq 7$ galaxies.

Caution is needed when performing these comparisons, given the considerable uncertainty of dust temperature in these objects, whose dust emission spectral region is not always well sampled. Dust temperature is degenerate with dust mass, thus large uncertainties are expected. This was recently pointed out by \citet{Choban2024a}, who conducted a series of zoom-in cosmological simulations. They concluded that although grain growth in the ISM is the dominant channel for dust production at $z \gtrsim 5$, their model still underpredicts observed dust masses at such high redshifts.


\begin{figure}
\centering
\includegraphics[width=0.75\columnwidth]{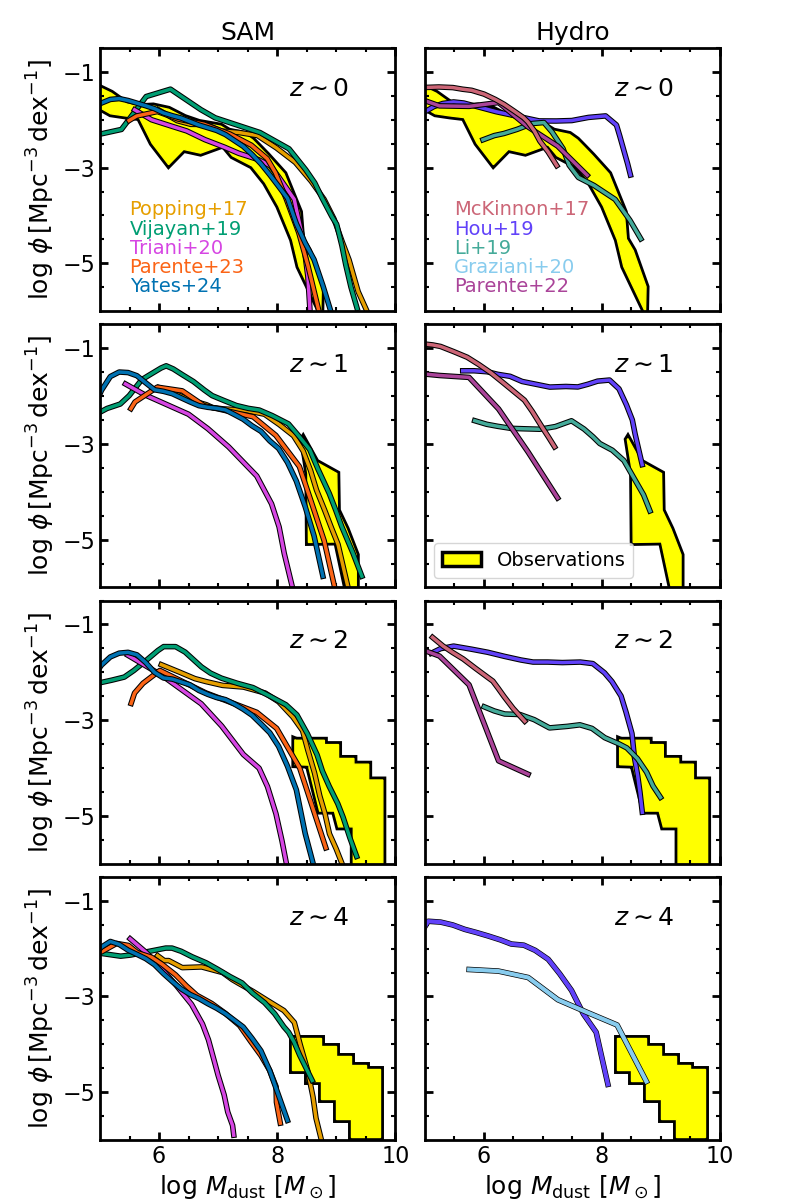}
\caption[DMF at $z\sim0,\,1,\,2,\,4$.]{DMF at $z\sim0,\,1,\,2,\,4$ as predicted by SAMs \citep[][left panels]{Popping2017,Vijayan2019,Triani2020,Parente2023,Yates2023}{}{} and hydrodynamic simulations \citep[][right panels]{McKinnon2017,Hou2019,Li2019,Graziani2020,Parente2022}{}{}. Yellow regions represent compilations of observational data taken from \citet{Vlahakis2005,Dunne2011,Beeston2018} at $z\sim 0$, \citet[][$z\sim 1$ and $\sim 2$]{Pozzi2020}, \citet[][$z \sim 2$]{Dunne2003}, \citet[][$z \sim 2$]{Berta25}, and \citet[][$z\sim 1$, $\sim 2$, and $\sim 4$]{Traina2024}.}
\label{fig:DMFrev}
\end{figure}


\begin{figure}
\centering
\includegraphics[width=\columnwidth]{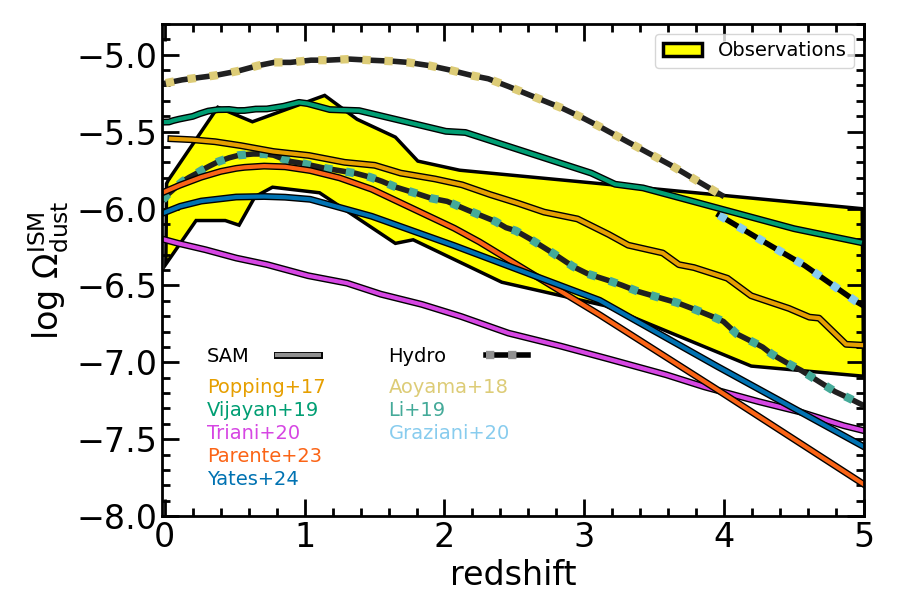}
\caption[$\Omega^{\rm ISM}_{\rm dust}$ as a function of redshift.]{Cosmic ISM dust parameter $\Omega^{\rm ISM}_{\rm dust}$ as a function of redshift. Model predictions from both SAMs \citep{Popping2017,Vijayan2019,Triani2020,Parente2023,Yates2023} and hydrodynamic simulations \citep{Aoyama2018,Li2019,Graziani2020} are shown as solid and dotted lines. The yellow shaded region includes a number of observational derivation of $\Omega^{\rm ISM}_{\rm dust}$ obtained by integrating the DMF \citep[][]{Vlahakis2005,Dunne2011,Beeston2018,Driver2018,Dudzeviciute2020,Pozzi2020,Traina2024,Berta25}.}
\label{fig:DMDrev}
\end{figure}

\subsection{The Dust Mass Function and Cosmic Dust Budget}

Simulating galaxy evolution in a cosmological context made it possible the comparison with IR-to-submm surveys, from which dust mass functions (DMFs) and information on the redshift evolution of cosmic dust abundance are inferred. Figure \ref{fig:DMFrev} reports DMF predictions by various SAMs and hydrodynamic simulations at $z\sim0,\,1,\,2,\,4$, compared with a compilation of observations (see caption for details). It is worth noting several aspects of this comparison. Firstly, the predicted DMF exhibits a Schechter-like shape in SAMs, whereas it appears more irregular in hydrodynamic simulations. It is crucial to bear in mind that SAMs typically yield a better agreement with the observed SMF than hydrodynamic simulations, largely due to the simplicity of running and tuning SAMs. Additionally, hydrodynamic simulations are often conducted on smaller volumes, which may result in inadequate sampling of the most massive objects. Both these factors are likely to influence the shape of the DMF.

The observed high-$z$ DMF is typically much more difficult to match with respect to the $z\sim 0$ one. \citet{Graziani2020} perform the best at $z\sim 4$, but a larger simulated volume (their box size is $30 h^{-1}\,{\rm cMpc}$) would be needed to better constrain the high mass end. At such high redshifts, all the other models underpredict the observed dust masses. A similar tendency is found at $z\sim2$. Here some SAMs \citep[][]{Popping2017,Vijayan2019}{}{} and hydrodynamic simulations \citep[][]{Hou2019,Li2019}{}{} broadly matches, still underestimating, the high mass end of the observed DMF. However, it is worth noting that predictions by these models typically worsen at $z \sim 0$. On the contrary, models that exhibit a good agreement with the $z \sim 0$ mass function, tend to (heavily) underestimate the observed DMF at $z \sim 1-2$ \citep[][]{Triani2020,McKinnon2017,Parente2022,Yates2023}{}{}. In conclusion, none of the shown models is able to reproduce both the high$-z$ and low$-z$ observed DMFs. 

A similar issue arises when examining the cosmic dust density $\rho_{\rm dust}$, or $\Omega_{\rm dust}$\footnote{The cosmic dust parameter is related to the cosmic dust density by $\Omega_{\rm dust}(z)=\frac{\rho_{\rm dust}(z)}{\rho_0}$, with $\rho_{\rm c,0}=2.775 \,h^2 \times 10^{11} \, M_\odot/{\rm Mpc}^3$ the critical density of the Universe today. {Note the common \textit{bad practice} in this definition, which is at odds with the standard meaning of the density parameters $\Omega$ in cosmology, where densities are normalized to the critical density at a given redshift $\Omega(z)=\frac{\rho(z)}{\rho_{\rm crit}(z)}$. }}. Here the focus is on galactic dust, hence $\Omega_{\rm dust}=\Omega^{\rm ISM}_{\rm dust}$. 
Predictions from both SAMs and hydrodynamic cosmological simulations are shown in Figure \ref{fig:DMDrev}. None of the models can adequately reproduce both the shape, characterized by a decline at $z \lesssim 1$, and the normalization of the observed $\Omega^{\rm ISM}_{\rm dust}$. In particular, as already noted for the DMF, models struggle to reproduce both the high$-$ and low$-z$ observed dust abundance. Reasons behind this discrepancy may be multiple and, importantly, not necessarily related to differences in the dust model. This has been noted recently by \citet{Parente2023}, who showed that a different prescription for SMBH growth (i.e. galaxy quenching) can improve the $z \lesssim 1$ drop prediction. Indeed, these simulations typically predict very different key quantities, such as SFR density or metals content, which profoundly affect the dust assembly, and consequently $\Omega^{\rm ISM}_{\rm dust}$.

As a final remark, it is important to highlight that $\Omega^{\rm ISM}_{\rm dust}$ may not serve as an ideal quantity for comparison with observations, where it is computed\footnote{At least in the examples considered here.} by integrating the observed DMF. The bulk of this integral ($\gtrsim 95\,\%$) arises from the contribution of galaxies with $M_{\rm dust}\lesssim 10^8 \, M_\odot$, a mass range that is largely unconstrained by high$-z$ observations, necessitating the use of DMF Shechter-like extrapolation to estimate its contribution. Such extrapolations can potentially lead to misleading comparisons. As an explicit example, take the DMF and $\Omega^{\rm ISM}_{\rm dust}$ at $z \sim 2$. While various models may align well with the observed $\Omega^{\rm ISM}_{\rm dust}$, none of them can accurately reproduce the high-mass end of the observed DMF. This highlights the importance of preferring comparisons based on the latter rather than on $\Omega^{\rm ISM}_{\rm dust}$.

\subsection{Dust on sub-grid scales}

Hydrodynamical simulations and spatially resolved SAMs allowed to study the dust distribution and properties on sub-galactic scales. Dust (as well as DTG and DTM) profiles have been investigated and compared with the few available observational results coming from nearby galaxies \citep[e.g.][]{Casasola2017,Relano2020,Chiang21}{}{}. 

Simulations recover slowly decreasing dust, DTG and DTM profiles, in broad agreement with observations \citep[][]{Granato2021, Yates2023}{}{}. However, the normalization often exhibits discrepancies, as indicated by offsets in various studies \citep[e.g.,][]{Aoyama2017, Parente2022, Choban2022, Romano2022}{}{}. The drop of DTM is particularly clear at the outer radii. \citet{Yates2023} noted that this is also influenced by the efficiency of dust destruction in the hot CGM, which supplies dust to the peripheral regions of the galactic disc where dust production is less efficient. \citet{Romano2022} highlighted the importance of dust diffusion in determining the profiles. The diffusion of large grains towards the outer regions of the galaxy makes the DTM profile shallower and drops at larger radii than those obtained when diffusion is neglected.

As for the grain size distribution, most numerical models predict a nearly flat or smoothly increasing profile of the ratio between small and large grains mass \citep{Aoyama2017, Granato2021,Parente2022, Romano2022, Parente2023}, as observations do \citep{Relano2020}. This suggests the balance between shattering and coagulation of grains on galactic scales.

\subsection{Dust on extra-galactic scales}


\begin{figure}
\centering
\includegraphics[width=\columnwidth]{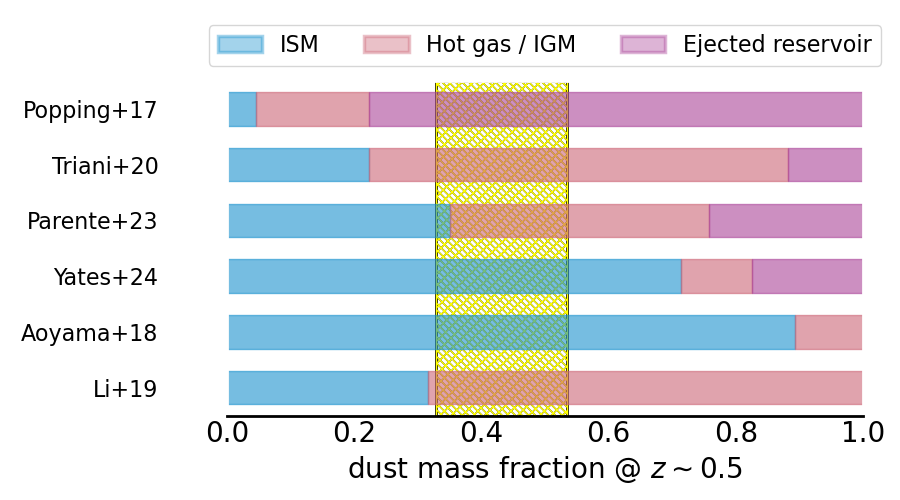}
\caption[Dust inside and outside galaxies at $z \sim 0.5$.]{Fraction of dust in and outside galaxies at $z \sim 0.5$ as predicted by different simulations (SAMs: \citealt{Popping2017}; \citealt{Triani2020}; \citealt{Parente2023}; \citealt{Yates2023}; hydrodyamic simulations: \citealt{Aoyama2018}; \citealt{Li2019}). Blue refers to dust in the ISM (or, equivalently, in galaxies), red to dust in the hot gas of SAMs or the IGM (i.e. outside galaxies) of hydrodynamic simulations, purple to dust in the ejected reservoir in SAM galaxies. The hatched yellow region represents a naive estimate of the ISM dust fraction at $z\sim 0.5$ by combining data from \citet{Menard2010,Menard2012} (for extra-galactic dust) and \citet{Driver2018,Pozzi2020} (for galactic dust).}
\label{fig:CGMdustrev}
\end{figure}

Cosmological simulations allow to follow the evolution of the hot CGM and IGM, hence the dust distribution in these extra-galactic environments. Figure \ref{fig:CGMdustrev} reports predictions from simulations concerning the fraction of dust inside and outside galaxies at $z \sim 0.5$. Dust inside galaxies is labeled as ISM dust, and it corresponds to dust in the cold gas of SAM galaxies and to dust bound in identified structures in hydrodynamic simulations. Dust outside galaxies is identified as dust not belonging to structures in hydrodynamic simulations (IGM), or belonging to both the hot and ejected reservoir in SAM galaxies. A naive estimation of the observed fraction of dust inside galaxies at $z\sim0.5$ has been obtained by combining results from \citet{Menard2010, Menard2012} for extra-galactic dust and from \citet{Driver2013, Driver2018} for galactic dust. It is reported for comparison. It should be kept in mind that this is an upper limit of the real galactic dust fraction, since a considerable part of the extra-galactic one may be missing.

The variation of model predictions is huge, with the extra-galactic dust fraction ranging from $\sim 10\% $ to $\sim 95\%$. In contrast, observations suggest a more balanced distribution between intra- and extragalactic dust (around $\sim 60 \%$ and $\sim 40 \%$, respectively). Since dust is only produced inside galaxies, key processes regulating extra-galactic dust are thermal sputtering in the hot medium and transport by galactic winds. Differences in the predictions of hydrodynamic simulations may stem from AGN feedback, which is implemented in \citet{Li2019} but not in \citet{Aoyama2018}. The latter tends to overestimate the stellar and dust mass function at $z=0$. In contrast, a process like AGN feedback may aid in reconciling these discrepancies with observations\footnote{This aspect has been addressed in \citet{Hou2019}. However, they do not specifically discuss the abundance of extra-galactic dust.}, also helping in ejecting dust and increasing the extra-galactic dust budget. 

As for SAMs, the ones predicting a larger amount of extra-galactic dust \citep{Popping2017,Triani2020} are the only ones having an implementation of AGN feedback capable of ejecting dust (and gas) from galaxies. However, the discrepancy with observations suggests the need to introduce a mechanism for dust destruction within such ejecta. The two remaining models \citep{Parente2023, Yates2023} are based on the same SAM \citep{Henriques2020}, but feature a factor of $\sim 2$ difference in the predictions of the ISM dust mass fraction. This is likely because \citet{Parente2023} adopts a smaller efficiency (by a factor of $10$) of the sputtering timescale\footnote{Also the hot gas density is computed differently (isotropic distribution is assumed in \citealt{Parente2023}, while a $\beta$ model is adopted in \citealt{Yates2023}).}. A reduced efficiency was previously suggested by simulations examining the dust content of the intra-cluster medium \citep{Gjergo2018, Vogelsberger2019}, and in this case allows the model to match the observed ISM dust fraction well.\\

Numerical simulations have also investigated extragalactic dust profiles, extending up to approximately $\simeq 1 \, {\rm Mpc}$ from the centers of galaxies \citep[][]{McKinnon2017, Aoyama2018, Parente2022}. Despite differences in the treatment of dust evolution and prescriptions for feedback and galactic winds, all of these simulations observe a similar slope of profiles ($\Sigma_{\rm dust} \propto r^{-0.8}$), broadly consistent with observations \citep{Menard2010}.


\section{Future prospects for dusty simulations}
\label{sec:futureprosp}
So far, simulations treating dust evolution have clarified many aspects of our understanding of the dust life cycle inside and outside galaxies. These simulations (and the next ones as well) have the unique opportunity to exploit dust to improve our galaxy evolution studies. This can be achieved in different ways.\\ 

\begin{itemize}
    \item \textbf{Realistic mock catalogs}
\end{itemize}

The first, and perhaps most obvious, improvement is to use dust to produce mock observational catalogs\footnote{In the literature, there are several examples of simulations that have been post-processed to produce mock catalogs \citep[e.g.][]{Lacey2016, Vijayan2021, Shen2022}{}{}. However, in these cases, the simulation does not track dust abundance consistently. Instead, it is typically linked to the ISM metals content by assuming a certain DTM ratio.} (as done in \citealt{Triani2023}; see also \citealt{Narayanan23} for PAHs spectral features). This method allows for a more coherent comparison with observations, as the observables become genuine predictions of the simulations, eliminating the need to rely on empirical relations to convert fluxes to physical quantities (e.g., dust mass from IR luminosity). This approach is particularly crucial at high redshifts, where observations are poor and the conversion from observed data to physical quantities is highly uncertain.\\

\noindent
The other intriguing way to exploit dust is to include it actively in the physics of galaxy evolution. Indeed, so far most of the simulations include dust as a passive ingredient of the gas phase of galaxy. A few notable exceptions include the incorporation of dust grain cooling in some hydrodynamic simulations \citep[][]{Vogelsberger2019, Granato2021}, as well as the work by \citet[][based on the \citealt{Bekki2013} dust model]{Osman20} who included Photoelectic Heating (PEH) in hydrodynamic simulations of MW like galaxies and studied its impact on the suppression of star formation. Another noteworthy case is the coupled modeling of dust and H$_2$ formation in cosmological simulations by \citet{RagoneFigueroa2024}, which marks a crucial step toward exploiting dust to enhance the still subgrid molecular cloud physics and star formation in galaxy evolution simulations. 
Some natural extensions of the dust models would include:

\begin{itemize}
    \item \textbf{Molecules formation on grains surfaces}
\end{itemize}

In a metal-enriched Universe, grains are the main channel for molecule formation, in particular H$_2$, the most abundant molecule in the Universe \citep[e.g.][]{Wakelam2017}{}{}. Since star formation in simulations is typically linked to H$_2$, incorporating this aspect significantly improves current simulations. This approach can be extended to other molecules, such as the well-observed CO, to provide a framework for simulating observable sub-mm lines.

\begin{itemize}
    \item \textbf{Radiative transfer}
\end{itemize}

Due to their interaction with energetic photons, dust grains significantly influence the physics of the ISM. For instance, dust shielding can affect the thermal properties of gas, thereby facilitating star formation. Also, in the presence of radiation fields from stellar sources or AGNs, radiation pressure must be considered, especially for the dusty medium. Radiation pressure on dusty gas can drive galactic outflows, which is crucial in the current galaxy evolution paradigm. Therefore, a proper dust-based model for this process is highly desirable. A first step in this direction, that is, the inclusion of radiation in dusty simulations, has been done in the \textsc{Thesan} project \citep{Kannan2022}.

\begin{itemize}
    \item \textbf{Planetary systems}
\end{itemize}

Coagulation of dust grains is the starting point for the formation of planetesimal (and then planets) in planetary discs \citep[see e.g.][and references therein]{Armitage2020}{}{}. Hence, dust abundance, properties and distribution within the ISM significantly influences planet formation. Although planetary scales are far below the resolution of any galaxy evolution simulation and their relevance to galaxy evolution simulations is likely negligible, dust can still serve as a foundation for simulations-based studies of galactic habitability.\\

While these goals are challenging and can be computationally demanding, they will be at some point achievable and would offer a more comprehensive understanding of the physical processes shaping the galaxies in our Universe, eventually enhancing our simulations' accuracy and predictive power.

\section*{\small{Acknowledgments}}

\small{Being this report a by-product of my PhD Thesis, it is essential to thank my supervisors: Cinthia Ragone-Figueroa, Gian Luigi Granato and Andrea Lapi. I want to further thank Gian Luigi for his many suggestions and for carefully reviewing this version of the report.\\
I also thank all the people with which I shared precious (dusty) discussions during these years. Among them: Laura Silva, Alessandro Bressan, Cedric G. Lacey, Francesco Calura, Robert M. Yates, Aswin P. Vijayan, Francesca Pozzi, Alberto Traina, Meriem Behiri, Roberta Tripodi.}

\newpage 
\newgeometry{margin=4cm, top=2cm, bottom=2cm} 
\begin{landscape}
\centering 
        {\small
        \begin{longtable}{ccccccccc}

         \hline
         \mycol{reference} & \mycol{scheme} & \mycol{chemical \\ composition} & \mycol{grain \\ sizes} & \mycol{dust \\ production} & \mycol{grains \\ accretion} & \mycol{SN \\ destruction} & \mycol{thermal \\ sputtering}\\
         \hline
         \hline
        
        \rowcolors{2}{lightgray}{white}
        
        
        \mycol{\citet{Bekki2013} \\ \citet{Bekki2015}} & \mycol{hydro} & \mycol{\barra} & \mycol{\barra} & \mycol{$\delta_{\rm fixed}$ \\ (SNII, SNIa, AGB)} & \mycol{fixed} & \mycol{fixed} & \mycol{\barra}  \\

        \hline

        
        \mycol{\citet{McKinnon2016} \\ \citet{McKinnon2017} \\ \citet{Vogelsberger2019}} & \mycol{hydro} & \mycol{Sil, Car} & \mycol{\barra} & \mycol{$\delta_{\rm fixed}$ \\ (SNII, SNIa, AGB)} & \mycol{\ngas, \Tgas, \Zgas} & \mycol{fixed} & \mycol{\ngas, \Tgas}  \\

        \hline

        
        \mycol{\citet{Popping2017}} & \mycol{SAM} & \mycol{Sil, Car} & \mycol{\barra} & \mycol{$\delta_{\rm fixed}$ \\ (SNII, SNIa, AGB)} & \mycol{\ngas, \hmol} & \mycol{chem} & \mycol{\ngas, \Tgas} \\

        \hline

        
        \mycol{\citet{Aoyama2017} \\ \citet{Aoyama2018}} & \mycol{hydro} & \mycol{\barra} & \mycol{2} & \mycol{$\delta_{\rm fixed}$ \\ (SNII)} & \mycol{\ngas, size, \hmol} & \ngas & \mycol{\ngas, size}\\

        \hline

        
        \mycol{\citet{Hou2017} \\ \citet{Hou2019}} & \mycol{hydro} & \mycol{Sil, Car} & \mycol{2} & \mycol{$\delta_{\rm fixed}$ \\ (SNII, SNIa, AGB)} & \mycol{\ngas, \Tgas \\ size, chem} & fixed & \mycol{\ngas, size} \\

        \hline

        
        \mycol{\citet{McKinnon2018}} & \mycol{hydro} & \mycol{\barra} & \mycol{$N$} & \mycol{dust yields tables \\ (SNII, SNIa, AGB)} & \mycol{\ngas, \Tgas, \Zgas, \\ size} & \mycol{\ngas, \Zgas, \\ size} & \mycol{\ngas, \Tgas, size}\\

        \hline

        
        \mycol{\citet{Gjergo2018}\\ \citet{Granato2021} \\ \citet{Parente2022}} & \mycol{hydro} & \mycol{Sil, Car} & \mycol{2} & \mycol{$\delta_{\rm fixed}$ \\ (SNII, SNIa, AGB)} & \mycol{\Zgas, \\ size, chem} & \mycol{fixed} & \mycol{\ngas, \Tgas, size} \\

        \hline

        
        \mycol{\citet{Vijayan2019}} & \mycol{SAM} & \mycol{Sil, Car,\\ SiC, Iron} & \mycol{\barra} & \mycol{dust yields tables (AGB) \\ $\delta_{\rm key\,element}$ (SNII, SNIa)} & \mycol{$M_{\rm dust}$, \\ chem} & \mycol{fixed} & \mycol{\barra} \\

        \hline

        
        \mycol{\citet{Li2019}} & \mycol{hydro} & \mycol{\barra} & \mycol{\barra} & \mycol{$\delta_{\rm fixed}$ (SNII, AGB)} & \mycol{\ngas, \Tgas, \Zgas} & \mycol{fixed} & \mycol{\ngas, \Tgas}\\

        \hline

        
        \mycol{\citet{Graziani2020}} & \mycol{hydro} & \mycol{\barra} & \mycol{\barra} & \mycol{dust yields tables \\ (PISN, SNII, AGB)} & \mycol{\ngas, \Tgas, \Zgas} & \mycol{fixed} & \mycol{\ngas, \Tgas}\\

        \hline

        
        \mycol{\citet{Triani2020}} & \mycol{SAM} & \mycol{\barra} & \mycol{\barra} & \mycol{$\delta_{\rm fixed}$ \\ (SNII, AGB)} & \mycol{\Zgas, \hmol} & \mycol{\Zgas} & \mycol{\ngas, \Tgas} \\

        \hline

        
        \mycol{\citet{Aoyama2020} \\ \citet{Romano2022}} & \mycol{hydro} & \mycol{\barra} & \mycol{$N$} & \mycol{$\delta_{\rm fixed}$ \\ (SNII, SNIa, AGB)} & \mycol{\ngas, \Tgas, \Zgas, \\ size} & \mycol{size} & \mycol{\ngas, \Tgas, size} \\

        \hline

        
        \mycol{\citet{Li2021}} & \mycol{hydro} & \mycol{\barra} & \mycol{$N$} & \mycol{$\delta_{\rm fixed}$ \\ (SNII, AGB)} & \mycol{\ngas, \Tgas, \Zgas, \\ size} & \mycol{size} & \mycol{\ngas, \Tgas, size} \\

        \hline

        
        \mycol{\citet{Dayal2022}} & \mycol{SAM} & \mycol{\barra} & \mycol{\barra} & \mycol{fixed mass \\ per SNII} & \mycol{\Zgas} & \mycol{fixed} & \mycol{\barra}  \\

        \hline

        
        \mycol{\citet{Choban2022}\\ \citet{Choban2024}} & \mycol{hydro} & \mycol{Sil, Car \\ SiC, Iron} & \mycol{\barra} & \mycol{dust yields tables (AGB),\\$\delta_{\rm key\,element}$ (SNII, SNIa) } & \mycol{\ngas, \Tgas, \hmol, \\ chem} & \mycol{\ngas} & \mycol{\ngas, \Tgas} \\

        \hline

        
        \mycol{\citet{Parente2023}} & \mycol{SAM} & \mycol{Sil, Car} & \mycol{2} & \mycol{$\delta_{\rm key\, element}$ (SNII, AGB)} & \mycol{\Zgas, \hmol, \\ chem, size} & \mycol{\Zgas} & \mycol{\ngas, \Tgas, size} \\

        \hline

        
        \mycol{\citet{Yates2023}} & \mycol{SAM} & \mycol{Sil, Car, \\ SiC, Iron} & \mycol{\barra} & \mycol{dust yields tables (AGB) \\ $\delta_{\rm key\, element}$ (SNII, SNIa)} & \mycol{$M_{\rm dust}$, \hmol, \\ chem} & \mycol{chem} & \mycol{\ngas, \Tgas} \\

        \hline

        
        \mycol{\citet{Dubois2024}} & \mycol{hydro} & \mycol{Sil, Car} & \mycol{2} & \mycol{$\delta_{\rm key\,element}$ \\ (SNII, SNIa, AGB) } & \mycol{\ngas, \Tgas, \Zgas, \\ size, chem} & \mycol{size, chem} & \mycol{\ngas, \Tgas \\ size, chem}  \\

        \hline

         \\
        \caption[Dust processes adopted in cosmological simulations published so far]{Schematic summary of the main processes ruling dust formation and evolution as implemented in different galaxy evolution simulations. Similar works, belonging to the same group, are organized together in the same raw. The description of the processes is a summary of the works, or refer to the most updated, fiducial version. 
        For each entry it is reported: \textbf{\textit{(i)} Main references.} \textbf{\textit{(ii)} Galaxy evolution scheme} SAM or hydrodynamic. \textbf{\textit{(iii)} Chemical composition of grains} Silicates (Sil), Carbonaceous (Car), SiC and Iron dust. Reported only when it is explicitly followed during both dust production and evolution. \textbf{\textit{(iv)} Number of grain sizes assumed} $2$ stands for the two-size approximation, $N$ is when a $N(>2)$--bins distribution is adopted. Note that the treatment of the grain size distribution implies the treatment of shattering and coagulation processes. \textbf{\textit{(v)} Dust production channels} $\delta_{\rm fixed}$ means that a fixed fraction of the metals (not necessarily the same for each metal) is assumed to condense into grains, $\delta_{\rm key\,element}$ refers to the adoption of the key element strategy to respect stechiometric ratios in the composition of produced dust, while dust yields tables means that pre-computed results are exploited. The stellar sources of dust are also reported. \textbf{\textit{(vi)} Physical dependences of the grains accretion process} These may include gas density ($n_{\rm gas}$), temperature ($T_{\rm gas}$), metals composition of the gas $Z_{\rm gas}$, molecular gas content (H$_2$), grain size and chemical properties of grains. \textbf{\textit{(vii)} Physical dependences of the SN destruction process} These may include $n_{\rm gas}$, $Z_{\rm gas}$, grain size and chemical properties of grains. These dependencies refer to the assumed mass swept of gas by a single SN and to the grain destruction efficiency. Fixed means that these quantities are constant in the simulation (but a dependence on the SN rate is always assumed). \textbf{\textit{(viii)} Physical dependences of the thermal sputtering process} These may include $n_{\rm gas}$, $T_{\rm gas}$, grains size and chemical composition.}
        \label{tab:subgriddust}
        \end{longtable}
        }
        


\end{landscape}
\restoregeometry

\bibliographystyle{mnras}  
\bibliography{bibThesis}  

\end{document}